\documentclass[aip,reprint]{revtex4-1}
\usepackage{array}
\usepackage{graphicx}
\usepackage[dvipsnames]{xcolor}
\usepackage{caption}
\usepackage{subcaption}
\usepackage{comment}
\usepackage[dvipsnames]{xcolor}
\usepackage{gensymb}
\usepackage{natbib}
\usepackage[%  
    colorlinks=true,
    pdfborder={0 0 0},
    linkcolor=blue,
    citecolor=blue
]{hyperref}

\draft

\begin{document}

\title{A novel method for thermal noise reduction, enabling measurements of broadband, low-amplitude electron temperature fluctuations using individual radiometer channels}

\author{C. Yoo}
\email[]{cyoo@mit.edu}
\affiliation{Plasma Science and Fusion Center, Massachusetts Institute of Technology, Cambridge, Massachusetts, MA 02139, United States of America}
\author{G. D. Conway}
\affiliation{Max Planck Institute for Plasma Physics, Boltzmannstr. 2, 85748 Garching, Germany}
\author{J. Schellpfeffer}
\affiliation{Plasma Science and Fusion Center, Massachusetts Institute of Technology, Cambridge, Massachusetts, MA 02139, United States of America}
\author{R. Bielajew}
\thanks{Present address: Rutherford Energy Ventures LP}
\affiliation{Plasma Science and Fusion Center, Massachusetts Institute of Technology, Cambridge, Massachusetts, MA 02139, United States of America}
\author{K. H\"ofler}
\affiliation{Max Planck Institute for Plasma Physics, Boltzmannstr. 2, 85748 Garching, Germany}
\affiliation{Technical University of Munich, TUM School of Natural Sciences, Physics Department, James-Franck-Str. 1, Garching, Germany.}
\author{D. J. Cruz-Zabala}
\affiliation{University of Seville, Seville, Spain}
\author{D. Cusick}
\thanks{Present address: Department of Physics, University of Oxford, Oxford, OX1 2JD, United Kingdom}
\affiliation{Plasma Science and Fusion Center, Massachusetts Institute of Technology, Cambridge, Massachusetts, MA 02139, United States of America}
\author{W. Burke}
\affiliation{Plasma Science and Fusion Center, Massachusetts Institute of Technology, Cambridge, Massachusetts, MA 02139, United States of America}
\author{B. Vanovac}
\affiliation{Plasma Science and Fusion Center, Massachusetts Institute of Technology, Cambridge, Massachusetts, MA 02139, United States of America}
\author{A. E. White}
\affiliation{Plasma Science and Fusion Center, Massachusetts Institute of Technology, Cambridge, Massachusetts, MA 02139, United States of America}
\author{the ASDEX Upgrade Team}
\thanks{See author list of H. Zohm \textit{et al.} 2024 \textit{Nucl. Fusion} \textbf{64} 112001}

\date{\today}

\begin{abstract}
A new analysis method has been developed for measurements of broadband, low-amplitude turbulent electron temperature fluctuations in fusion plasmas using individual radiometer channels of a Correlation Electron Cyclotron Emission (CECE) diagnostic. This method takes advantage of differences in the correlation time of thermal noise compared to the correlation time of plasma fluctuations in fusion reactors. The validation of this single-channel method is demonstrated using comparisons with the standard dual-channel radiometer spectral decorrelation method for measurements of turbulent electron temperature fluctuations in the core and edge of low confinement (L), improved confinement (I), and high confinement (H)-mode plasmas at the ASDEX Upgrade tokamak.
\end{abstract}

\maketitle

\section{Introduction}
\label{sec:Intro}

Turbulent fluctuations drive the majority of energy, particle, and momentum transport in tokamak plasmas.\cite{horton_drift_1999} Measurements of these fluctuations are essential for guiding the development of turbulent transport models used to design future fusion power plants. A variety of diagnostics are employed for fluctuation measurements of turbulent transport-relevant plasma parameters, including the electron temperature and density as well as the electric potential. The main diagnostic for measurements of electron temperature fluctuations throughout the magnetically-confined plasma volume is the Correlation Electron Cyclotron Emission (CECE) diagnostic.\cite{creely_correlation_2018, yoo_christian_study_2025} The CECE diagnostic is a specialized variant of the Electron Cyclotron Emission (ECE) diagnostic and is designed to enhance sensitivity to turbulent electron temperature fluctuations.

ECE diagnostics are well-suited to the measurement of electron temperature profiles. When the optical depth is high enough ($\tau \gg 1$) for the plasma to be considered a blackbody emitter of radiation, the intensity of ECE measured by each radiometer channel corresponds to the electron temperature. \cite{ian_h_hutchinson_principles_nodate} In addition, ECE diagnostics can be used to measure large-amplitude fluctuations with signal bandwidths ($B_{\rm sig}$) narrower than the diagnostic video bandwidth ($B_{\rm vid}$). \cite{udintsev_first_2006} Measurements of large-amplitude, narrowband magnetohydrodynamic (MHD) modes have been carried out previously using individual ECE and ECE Imaging (ECEI) radiometers (see, e.g., Refs. \citenum{udintsev_first_2006, white_experimental_2008, wang_millimeter-wave_2017, kohn-seemann_electron_2025}). Additionally, large-amplitude electron temperature fluctuations associated with the Weakly Coherent Mode (WCM) in the edge of an improved confinement (I-)mode plasma in Alcator C-Mod were measured using a single radiometer channel. \cite{white_electron_2011} However, each individual radiometer channel in an ECE diagnostic is sensitive to thermal noise that results from the wave-like nature of the measured photons and, in particular, the beating of randomly phased waves emitted by the plasma, which is an incoherent source of radiation.\cite{udintsev_first_2006, cima_correlation_1994, watts_review_2007} Thermal noise is generally multiple times larger in amplitude compared to broadband turbulent fluctuations such as drift-wave turbulence, which is prevalent in the core of tokamak plasmas. \cite{udintsev_first_2006} As a result, thermal noise has historically prevented accurate measurements of broadband, low-amplitude electron temperature fluctuations using individual radiometer channels in ECE diagnostics. 

\section{Background}
\label{sec:Background}

The radiometer sensitivity limit below which the root-mean-square (RMS) amplitude of electron temperature fluctuations ($\delta T_e/T_e$) is indistinguishable from thermal noise is given by Refs. \citenum{cima_correlation_1994, watts_review_2007} as:
\begin{equation} \label{Eqn: single_channel_limit}
    \frac{\delta T_e}{T_e} \bigg |_{\rm min} = \sqrt{\frac{2B_{\rm sig}}{B_{\rm IF}}}
\end{equation}
where $B_{\rm sig} \leq B_{\rm vid}$ and $B_{\rm IF}$ is the bandwidth of the radiometer intermediate frequency (IF) filter. Broadband, low-amplitude fluctuation levels in tokamak plasmas are frequently below this threshold, limiting their measurement with individual ECE radiometers. Although thermal noise also results from the finite temperature of the radiometer itself, with noise temperatures on the order of eV, this is dominated by the thermal noise from hot fusion plasmas in which the plasma temperature is on the order of keV.\cite{white_experimental_2008}

Prior to the development of CECE, turbulent electron temperature fluctuations in the Princeton Large Tokamak (PLT) were initially reported to have been measured using an individual ECE radiometer in 1977.\cite{arunasalam_turbulent_1977} However, later work on the TFR tokamak in 1981 showed that the apparent fluctuations measured by individual ECE radiometers were likely attributable to thermal noise and were not necessarily associated with actual turbulent electron temperature fluctuations.\cite{cavallo_measurement_1981}

The development of the CECE diagnostic has enabled accurate measurements of these broadband, low-amplitude fluctuations. The primary purpose of CECE is to improve the signal-to-noise ratio (SNR) of radiometer measurements and to enhance diagnostic sensitivity to electron temperature fluctuations in fusion plasmas. Multiple techniques exist for the implementation of CECE, including spectral decorrelation, detector decorrelation, spatial decorrelation, and autocorrelation with thermal noise subtraction.\cite{watts_review_2007} The first three of these techniques are based on correlating the signals from pairs of individual radiometer channels in order to remove uncorrelated thermal noise while retaining correlated turbulent fluctuations. In spectral decorrelation, the channel pairs measure in different frequency bands that map to volumes of plasma within the same radial correlation length of the turbulence but between which the thermal noise is uncorrelated. In detector decorrelation (also known as crossed-sightline decorrelation), two separate, non-overlapping optical systems observing the same plasma volume are used. In spatial decorrelation, two separate optical systems are used to measure at toroidally (or poloidally) separate plasma volumes. In each case, the thermal noise measured by the two separate systems is uncorrelated. Correlation analysis and ensemble averaging of these two distinct signals therefore reduce the thermal noise level and improve the SNR of the correlated signal. The first successful CECE measurements of core plasma electron temperature fluctuations were performed on the W7-AS stellarator in 1994 using the detector (crossed-sightline) decorrelation technique.\cite{sattler_experimental_1994} The first spectral decorrelation CECE measurements on a tokamak were carried out in the TEXT-U tokamak in 1995.\cite{cima_core_1995} 

In the method of autocorrelation with thermal noise subtraction, as stated in Ref. \citenum{watts_review_2007}, the thermal noise could (in principle) be removed by filtering out a sharp peak in the autocorrelation function. Since the autocorrelation function is related to the auto-power spectral density by the Wiener-Khinchine theorem (Ref. \citenum{bendat_random_2010}), an equivalent operation would be to subtract out a background level from the auto-power spectral density function. However, the auto-correlation with background subtraction method requires that the turbulent signal is sufficiently strong (i.e., has a sufficient SNR) to facilitate distinguishing the signal from the thermal noise in the measurement and enable the thermal noise floor to be identified accurately and thus subtracted out. For this same reason, this method requires that $B_{\rm sig} \ll B_{\rm vid}$. A method of this type was proposed in Ref. \citenum{thomas_autocorrelation_1990} in 1990 based on exploiting differences in the thermal noise correlation time and the turbulence correlation time by employing correlations between an ECE signal and a time-delayed version of itself, where the fixed time delay is achieved using hardware. The proposed method required two radiometer channels, using identical IF filters and detectors, for each signal and time-delayed signal pair. However, it was subsequently reported that this method would not work.\cite{watts_review_2007} 

Compared to the method of autocorrelation with background subtraction, two-channel CECE analysis methods such as spectral decorrelation can make measurements below the single-channel radiometer thermal noise floor and improve the SNR via ensemble averaging as indicated by the following equation for the spectral decorrelation sensitivity limit per Ref. \citenum{creely_correlation_2018}:
\begin{equation} \label{Eqn: two_channel_limit}
    \frac{\delta T_e}{T_e} \bigg |_{\rm min} = \sqrt{\frac{2}{\sqrt{N}} \frac{B_{\rm sig}}{B_{\rm IF}} \sqrt{\frac{f_s}{2 B_{\rm sig}}}}
\end{equation}
where $f_s$ is the sample rate, $N$ represents the number of independent samples and is given by $N = f_{s}\Delta t$, $\Delta t$ is the total duration of time used in the correlation analysis, and $B_{\rm sig} \leq B_{\rm vid}$. For this reason, measurements using two-channel CECE analysis are more sensitive to low-amplitude fluctuations compared to the single-channel autocorrelation with background subtraction method described in Ref. \citenum{watts_review_2007}.

\begin{figure*}
     \centering
     \begin{subfigure}[b]{0.8\textwidth}
         \centering
         \includegraphics[width=\textwidth]{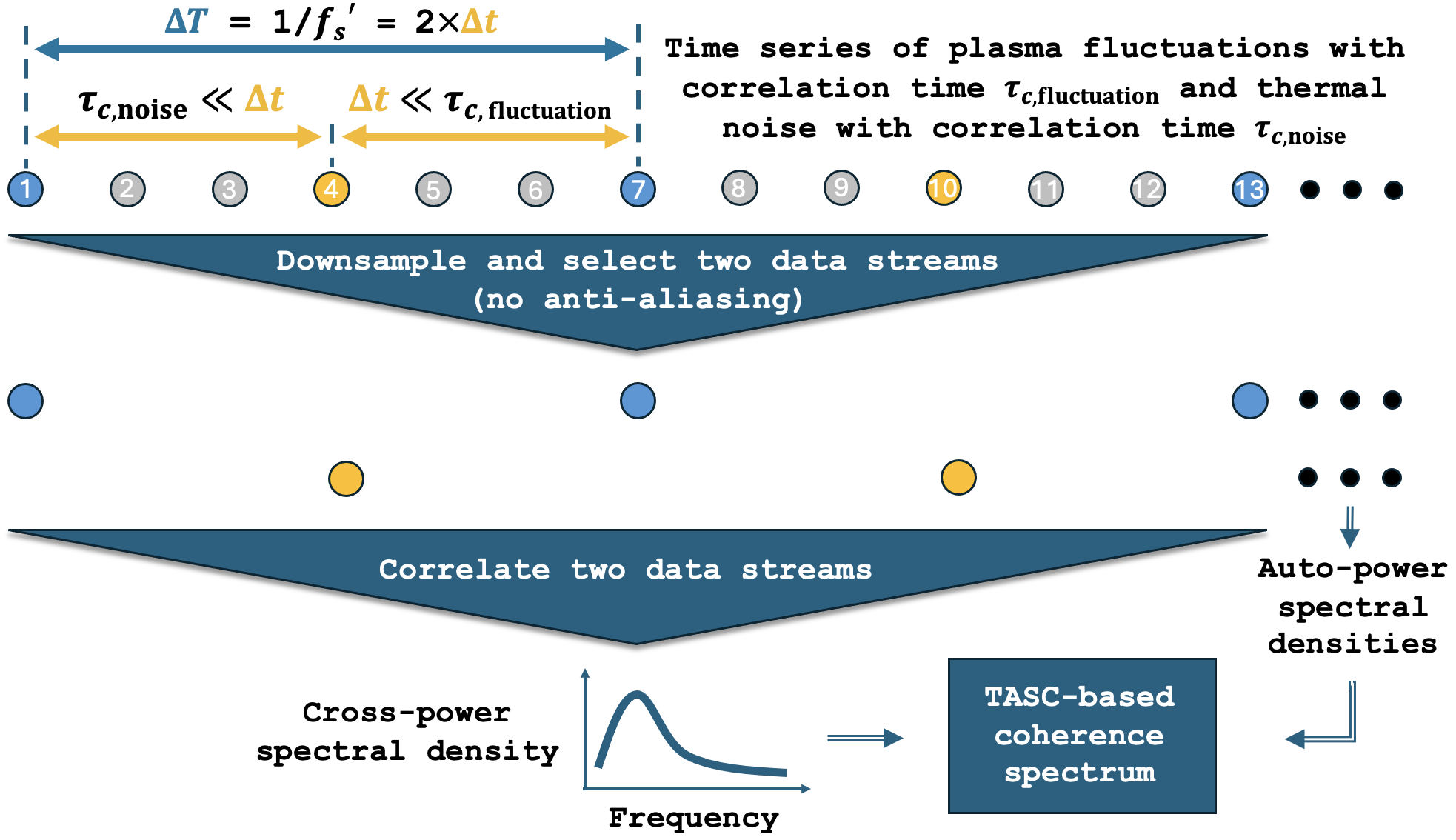}
     \end{subfigure}
     
    \caption{Overview of the Time Alternating Self-Correlation (TASC) method. In this example, the original time series from a single radiometer channel with sample rate $f_{s}$ is downsampled into six data streams (without anti-aliasing), each with sample rate $f_{s}^{'} = f_{s}/6$ and half-period $\Delta t$. The amount of downsampling is chosen so that $\Delta t$ is much longer than the correlation time of the thermal noise but much shorter than the correlation time of the plasma fluctuations. The first and fourth data streams are then correlated to produce a cross-power spectral density. A TASC-based coherence spectrum is calculated using the downsampled auto-power and cross-power spectral densities.}
    \label{fig:TASC_overview}
\end{figure*}

Although CECE based on two-channel methods such as spectral decorrelation continues to yield tremendous value in enabling cutting-edge measurements of turbulent electron temperature fluctuations, the prospect of obtaining an equivalent sensitivity to low-amplitude fluctuations using only a single radiometer channel is appealing for a number of reasons. For example, a successful single-channel CECE method would enable a fixed number of radiometer channels to be spread out over a wider radial range of the plasma compared to standard two-channel CECE, increasing diagnostic coverage of the plasma. Conversely, for a fixed radial range of diagnostic coverage, the implementation of a successful single-channel method would enable enhanced diagnostic spatial resolution since turbulence measurements could be obtained from both single channels as well as pairs of channels measuring in non-overlapping frequency bands. Additionally, the radial resolution associated with single-channel method measurements could be further improved since the IF filters could be configured to overlap in frequency space without leading to the increase in thermal noise that would otherwise degrade dual-channel spectral decorrelation CECE measurements. 

This paper re-examines this nearly 50-year-old challenge and presents a new method to facilitate broadband, low-amplitude fluctuation measurements using only a single ECE radiometer channel. This new single-channel method takes advantage of differences in the correlation time of thermal noise and the correlation time of plasma fluctuations, as was the case for the method proposed in Ref. \citenum{thomas_autocorrelation_1990}. However, instead of performing correlations between a signal and a time-delayed version of that signal as proposed in Ref. \citenum{thomas_autocorrelation_1990} (which required two channels and a hardware-based time delay and was reported to be ineffective in Ref. \citenum{watts_review_2007}), the method detailed in this work utilizes correlations between two downsampled versions of the same signal. As with standard dual-channel CECE methods, time averaging is employed to lower the thermal noise floor and enhance the SNR.

The remainder of this paper is organized as follows. Section \ref{sec:methods} provides an overview of the new single-channel method. Section \ref{sec:results} validates the method by comparing results from the new single-channel method to those from the conventional dual-channel spectral decorrelation method using measurements from the ASDEX Upgrade (AUG) tokamak. Section \ref{sec:discussion} gives a discussion of the results and Section \ref{sec:Conclusion} concludes the paper. Additional details relevant to the implementation of this new method are provided in the \hyperref[sec: Appendix]{Appendix}. Note that a version of the material in this paper previously appeared in chapters 4 and 7 in [C. Yoo, ``Study of turbulent electron temperature fluctuations and their cross-phase angles with electron density fluctuations at the ASDEX Upgrade tokamak", PhD thesis, Massachusetts Institute of Technology, submitted July 2025] (Ref. \citenum{yoo_christian_study_2025}).

\section{Methods}
\label{sec:methods}

The method presented here involves the correlation of two downsampled data streams selected from the same original time series. Datapoints within the two data streams are alternating in time and temporally separated from each other so that the plasma fluctuations measured by each data stream are correlated but the thermal noise is uncorrelated and can therefore be reduced via correlation analysis. While the individual steps presented in this section use standard signal processing and analysis techniques, including downsampling (Refs. \citenum{li_tan_digital_2014, conway_short_nodate}) and Fourier transforms, the overall method presented here appears to be new to plasma physics diagnostics and may also be novel within the fields of signal processing and analysis, to the best of the authors' knowledge. In particular, the principal features of this new method are not fully encapsulated by either standard auto-correlation or cross-correlation techniques (or, likewise, by auto-power or cross-power techniques). Therefore, this paper refers to the new method as the Time Alternating Self-Correlation (TASC) method. Details of the method are as follows and an overview of this method is shown in Figure \ref{fig:TASC_overview}.

A fundamental requirement of the TASC method is that the thermal noise in the radiometer measurements has a correlation time ($\tau_{\rm c,noise}$) that is much shorter than the plasma fluctuation correlation time ($\tau_{\rm c,fluctuation}$): $\tau_{\rm c,noise} \ll \tau_{\rm c,fluctuation}$. This requirement is fulfilled by appropriate hardware design of the radiometer, as explained here. The correlation times of thermal noise and plasma fluctuations are defined as follows per Refs. \citenum{udintsev_first_2006, white_electron_2011, watts_comparison_2004}: 
\begin{equation} \label{Eqn: noise_correlation_time}
    \tau_{\rm c,noise} = \sqrt{\frac{\ln 2}{\pi}}\frac{1}{B_{\rm vid}} \approx \frac{1}{2.13 \, B_{\rm vid}}
\end{equation}
\begin{equation} \label{Eqn: signal_correlation_time}
    \tau_{\rm c,fluctuation} = \sqrt{\frac{\ln 2}{\pi}}\frac{1}{B_{\rm sig}} \approx \frac{1}{2.13 \,B_{\rm sig}}
\end{equation}
Therefore, a large $B_{\rm vid}$ yields a short thermal noise correlation time while $B_{\rm sig} < B_{\rm vid}$ yields a comparatively longer plasma fluctuation correlation time. The AUG CECE system is designed with $B_{\rm vid}$ = 1 MHz (Ref. \citenum{creely_correlation_2018}), yielding a broadband thermal noise correlation time of about 0.47 microseconds. In contrast, in tokamak plasmas, where the signal bandwidths of turbulent electron temperature fluctuations are generally on the order of 10's of kHz to over 100 kHz, the correlation time of turbulent fluctuations can be almost one to two orders of magnitude longer (e.g., 5 to 50 microseconds), thereby fulfilling this first condition.

The TASC method involves downsampling a radiometer signal to separate the original steady-state time series with sample rate $f_{\rm s}$ into multiple data streams with a lower sample rate $f_{\rm s}^\prime$. This process results in $N = f_{\rm s}/f_{\rm s}^\prime$ downsampled data streams, where $N$ is an even integer greater than 1. In the example shown in Figure \ref{fig:TASC_overview}, the original time series is downsampled by a factor of 6 into 6 data streams, i.e., $N = 6$. The downsample factor is chosen such that two downsampled data streams with the following properties can be selected:

\begin{itemize}
\item First, the downsampled data stream half-period ($\Delta t = 1/(2f_{\rm s}^\prime)$, which is also the temporal separation between the two selected downsampled data streams) must be much longer than the thermal noise correlation time, i.e., $\tau_{\rm c,noise} \ll \Delta t$ $\rightarrow$  $f_{\rm s}^\prime \ll B_{\rm vid}$.
\item Second, the downsampled data stream half-period must be much shorter than the turbulence correlation time, i.e., $\Delta t \ll \tau_{\rm c,signal}$ $\rightarrow$ $B_{\rm sig} \ll f_{\rm s}^\prime$. The more restrictive condition $B_{\rm sig} \ll f_{\rm s}^\prime/2$, where $f_{\rm s}^\prime/2$ is the downsampled data stream Nyquist frequency, results from the necessity to resolve $B_{\rm sig}$ within the resulting frequency spectra.
\item Third, in cases where $B_{\rm sig} \ll f_{\rm s}^\prime/2$ but $B_{\rm sig}$ is nevertheless located at high frequencies, the downsampled data stream sample rate ($f_{\rm s}^\prime$) must be large enough such that $f_{\rm sig,max} < f_{\rm s}^\prime / 2$, where $f_{\rm sig,max}$ is the high frequency limit of $B_{\rm sig}$, so that the signal bandwidth can be accurately identified within the resulting frequency spectra.
\end{itemize}

After downsampling has been conducted, standard signal processing and analysis methods are applied, treating the two downsampled data streams (labeled here as $x$ and $y$) as two distinct time series. In particular, spectral estimates including the downsampled signals' single-sided auto-power ($G_{xx}$, $G_{yy}$) and cross-power ($G_{xy}$) spectral densities are calculated, where $G_{xy} = 2F^{*}_x(f)F_{y}(f)$, $F$ denotes a windowed and ensembled-averaged Fourier transform, and $^*$ indicates a complex conjugate.\cite{creely_correlation_2018} The TASC-based coherence spectrum is then calculated as follows using the standard equation for a coherence spectrum (e.g., Refs. \citenum{bendat_random_2010, creely_correlation_2018}) but, importantly, using the auto-power and cross-power spectral densities of the two downsampled data streams:
\begin{equation} \label{Eqn: coherence}
    \gamma_{\rm TASC}(f) = \frac{G_{xy}(f)}{\sqrt{G_{xx}(f)G_{yy}(f)}}
\end{equation}

The thermal noise floor is reduced by correlation analysis with ensemble averaging. Increasing the amount of downsampling and therefore the temporal separation of the data streams is analogous to increasing the degree of IF filter separation in the dual-channel spectral decorrelation technique, which is illustrated in Figure 3.20 in Ref. \citenum{white_experimental_2008}. The chosen downsampling factor should yield a TASC-based coherence level in the high frequency region above the signal bandwidth of the plasma fluctuations that is approximately equal to the coherence sensitivity limit given by the standard deviation predicted by Eqn. 5 in Ref. \citenum{creely_correlation_2018}. A coherence value of zero is applied to Ref. \citenum{creely_correlation_2018} Eqn. 5 to obtain the theoretical sensitivity level for two completely incoherent signals, which is expected in the high frequency region where only uncorrelated thermal noise should be present. The one-standard deviation coherence sensitivity limit is then: 
\begin{equation} \label{Eqn: coherence_sensitivity_limit}
    \sigma_{\rm \gamma} = \sqrt{\frac{1}{2n_d}}
\end{equation}
where $n_d$ is the number of independent data segments used in the ensemble average for calculations of the coherence. The TASC-based coherence spectrum can then be used to calculate $\delta T_e/T_e$ using Eqn. 1 in Ref. \citenum{molina_cabrera_isotope_2023}.

\begin{figure*}
     \centering
     \begin{subfigure}[b]{0.65\textwidth}
         \centering
         \includegraphics[width=\textwidth]{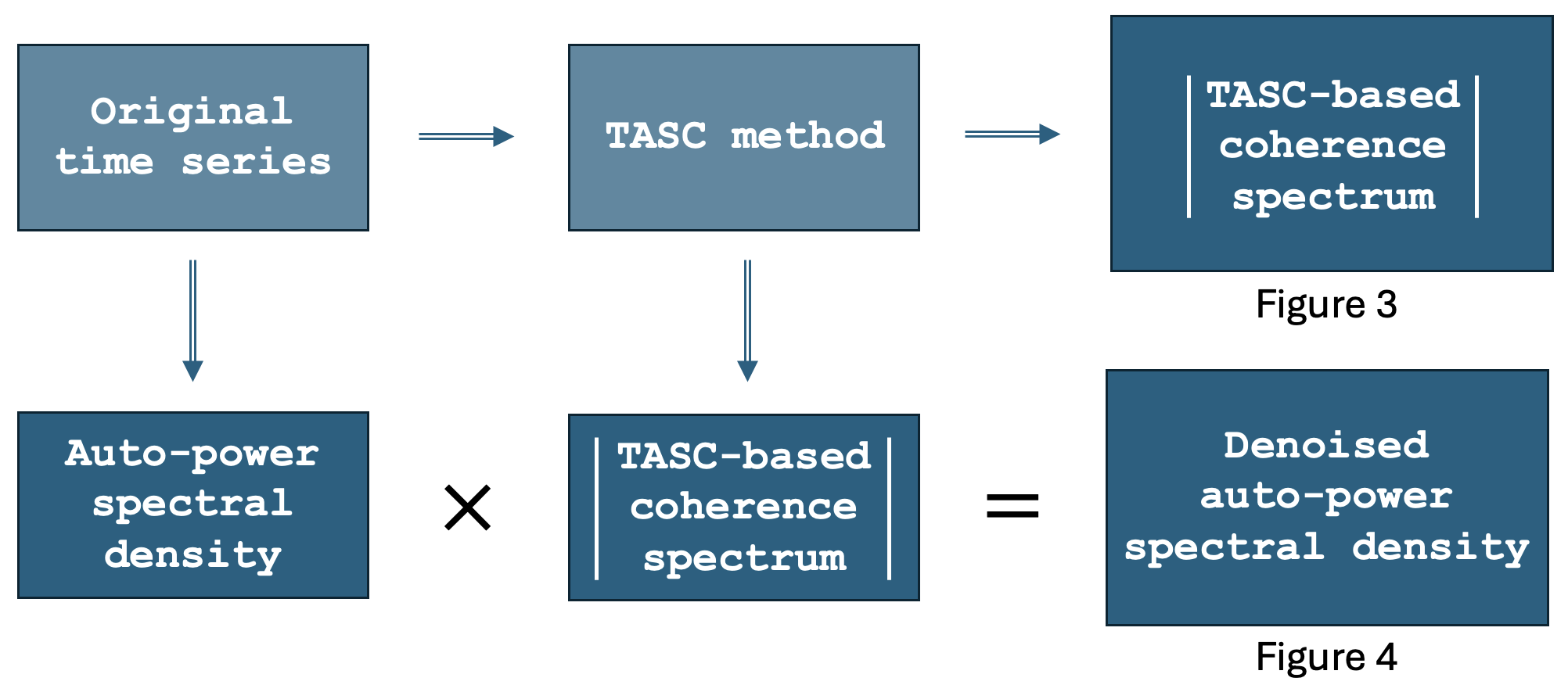}
     \end{subfigure}
     
    \caption{The auto-power spectral density calculated from the original time series from a single radiometer channel can be multiplied by the magnitude of the coherence spectrum calculated using the TASC method. This operation results in the denoised auto-power spectral density, in which the thermal noise floor is reduced and which is therefore characterized by an enhanced SNR compared to the auto-power spectral density of the original time series.}
    \label{fig:TASC_flowchart}
\end{figure*}

In addition, once the TASC-based coherence spectrum is calculated, its magnitude can be multiplied by the auto-power spectral density calculated from the original time series, as illustrated in Figure \ref{fig:TASC_flowchart}. This operation produces a denoised auto-power spectral density, in which the thermal noise floor is reduced compared to the auto-power spectral density of the original time series. The SNR of the auto-power spectral density is thus improved.

While hardware-based low-pass filtering of the radiometer signals should be carried out below the digitizer (non-downsampled) Nyquist frequency to prevent aliasing during digitization, it is important to note that the TASC method does not work if low-pass filtering below the downsampled Nyquist frequency is conducted during the downsampling process. The reason is that low-pass filtering increases the correlation time of the thermal noise in the downsampled data streams, as indicated by Eqn. \ref{Eqn: noise_correlation_time}. In particular, if a digital low-pass filter is applied prior to downsampling, such that the resulting video bandwidth $B_{\rm vid} \leq f_{s}^{\prime}/2$, then by Eqn. \ref{Eqn: noise_correlation_time} this results in $\tau_{\rm c,noise}^{'} > 1/f_{s}^{\prime} = 2 \Delta t$, which violates the first criterion for selecting a downsample factor. Therefore, the aliasing of thermal noise is unavoidable during downsampling. On the other hand, it is important to minimize electronics noise all the way up to the digitizer Nyquist frequency in order to prevent the aliasing of this noise during downsampling. It is possible to identify narrowband aliased electronics noise peaks that result from downsampling by comparing the auto-power spectral density calculated from the original (non-downsampled) time series to the auto-power spectral densities calculated from the downsampled data streams. However, the aliasing of broadband, non-thermal electronics noise is difficult to distinguish from broadband plasma fluctuations at low frequencies.

\section{Results}
\label{sec:results}

This section presents a validation of the TASC method through comparisons of TASC-based fluctuation spectra to those produced using the dual-channel spectral decorrelation method with AUG CECE diagnostic data. The CECE diagnostic at AUG has a sampling rate of 4 MHz and a video bandwidth of 1 MHz, set by (hardware-based) 4-pole Butterworth anti-aliasing filters prior to digitization. The 1 MHz video bandwidth of the diagnostic yields a thermal noise correlation time of approximately 0.47 microseconds. IF filter bandwidths are 200 MHz, with a spacing of 250 MHz between the center frequencies of frequency-adjacent IF filters.

\begin{figure*}
    \centering
    \begin{subfigure}[b]{0.55\textwidth}
         \centering
         \includegraphics[width=\textwidth]{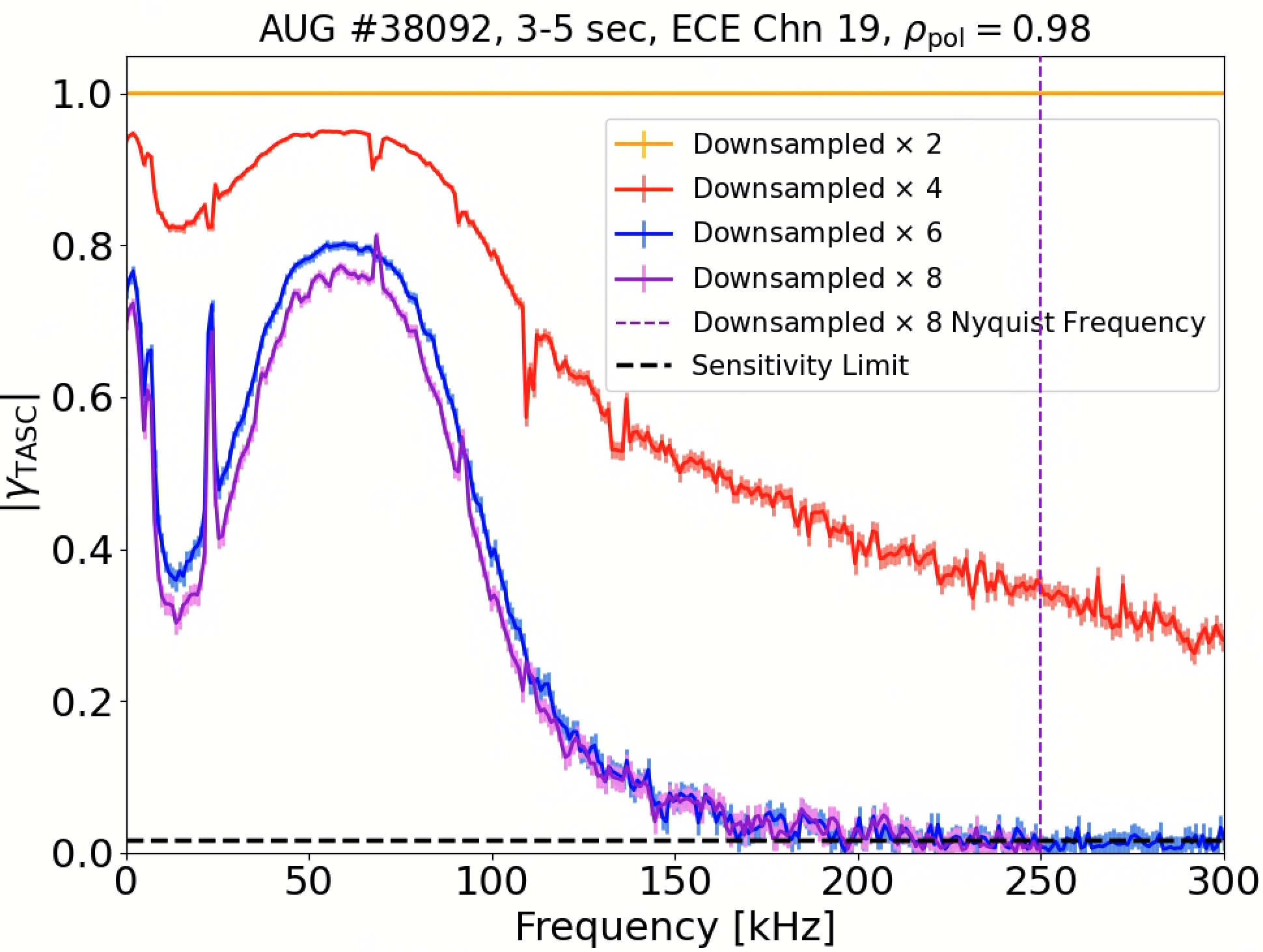} 
    \end{subfigure}
    \caption{The magnitudes of the TASC-based coherence spectra of the I-mode WCM at $\rho_{\rm pol} = 0.98$ for downsample factors 2, 4, 6, and 8, where $\rho_{\rm pol}$ is defined as the square root of the normalized poloidal magnetic flux. At high frequency (above 200 kHz, corresponding to the upper end of the frequency range of the plasma fluctuations), the coherence drops to the sensitivity limit as the downsample factor is increased to 6. The sensitivity limit is calculated the same way as for dual-channel CECE. Narrowband peaks are the result of electronics noise.}
    \label{fig:Figure_downsample_factor_scan}
\end{figure*}

Figure \ref{fig:Figure_downsample_factor_scan} shows the magnitude of the TASC-based coherence spectra resulting from a scan of downsample factors using measurements of electron temperature fluctuations associated with the WCM in the edge of an I-mode plasma (discharge 38092) at AUG as reported in Ref. \citenum{bielajew_edge_2022}.  The TASC method detailed in Section \ref{sec:methods} and illustrated in Figure \ref{fig:TASC_overview} was applied to single-channel measurements of the WCM for downsample factors of 2, 4, 6, and 8. The auto-power and cross-power spectral densities of the downsampled signals were calculated using Welch's method for fast Fourier transforms (FFTs) with 50\% overlapping Hamming windows and approximately 1 ms data segments, yielding $\sim$1 kHz frequency resolution.\cite{welch_use_1967, bendat_random_2010} 

Increasing the downsample factor from 2 up to 6 is shown to improve the single-channel SNR in Figure \ref{fig:Figure_downsample_factor_scan}, distilling the broadband plasma fluctuations from the thermal noise, which is reduced by the correlation process. Note that the narrowband peaks are the result of electronics noise. The radiometer time series were digitally low-pass filtered using 4-pole Butterworth filters at 0.9 MHz to reduce the aliasing of broadband, high frequency electronics noise during downsampling. Further details on the rationale for implementing this digital low-pass filter are given in the \hyperref[sec: Appendix]{Appendix}. Downsampling by a factor of 6 yields a maximum broadband coherence level of the WCM of about 0.8 at 60 kHz while reducing the coherence at high frequencies (greater than about 200 kHz) above the signal bandwidth of the WCM to the sensitivity limit given by Eqn. \ref{Eqn: coherence_sensitivity_limit}, which is the same sensitivity limit used for dual-channel CECE. Downsampling by a factor of 8, however, reduces the coherence level associated with the plasma fluctuations at frequencies less than about 110 kHz without a statistically significant reduction of thermal noise and is therefore undesired. Since the thermal noise floor has already been reduced to the level of the sensitivity limit using a downsample factor of 6, this downsample factor is determined to be optimal for this measurement.

\begin{figure*}
    \centering
    \begin{subfigure}[b]{0.55\textwidth}
         \centering
         \includegraphics[width=\textwidth]{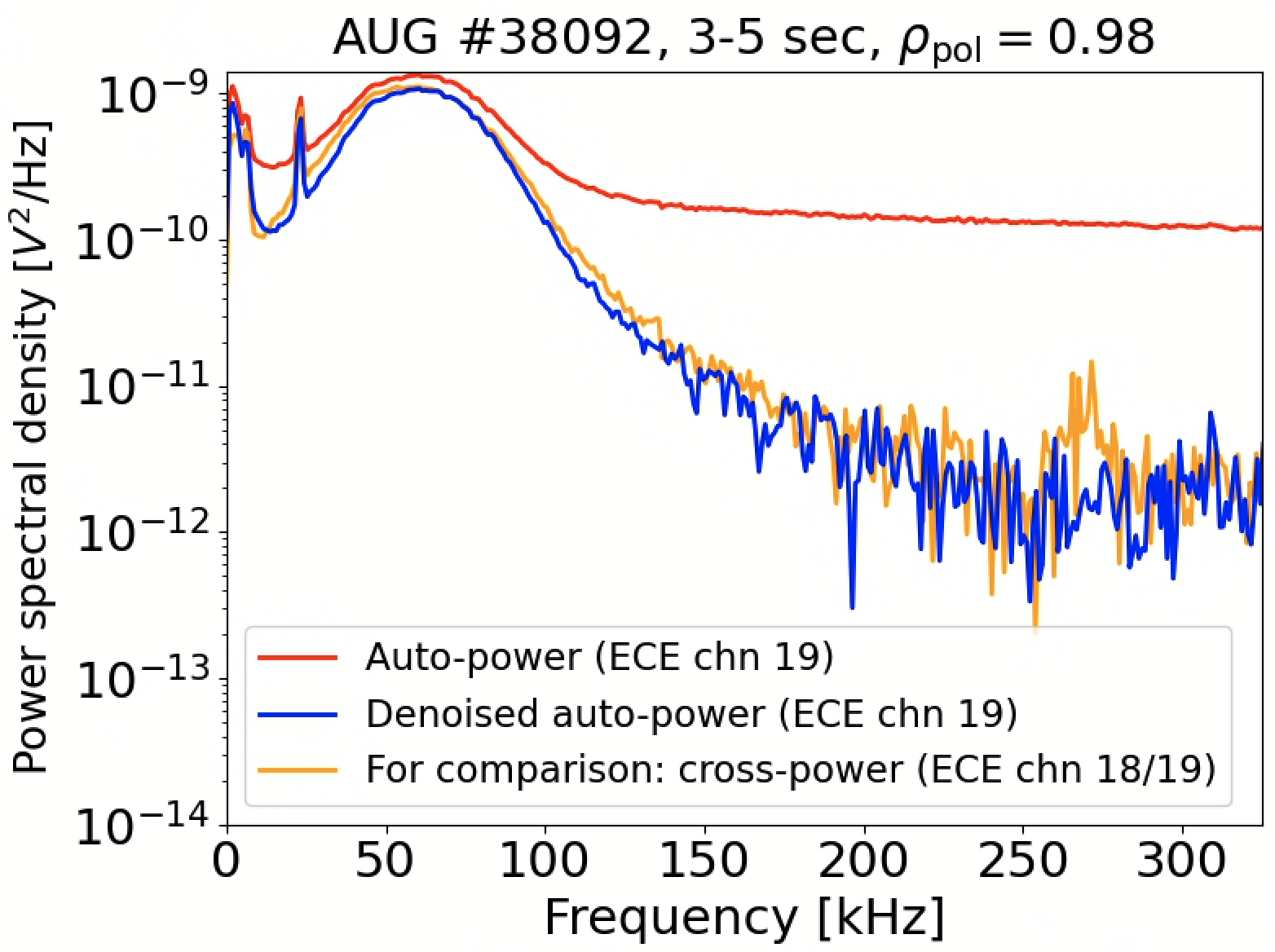} 
    \end{subfigure}
    \caption{The denoised auto-power spectral density exhibits a greater SNR compared to the auto-power spectral density of the original time series for measurements of the I-mode WCM at $\rho_{\rm pol} = 0.98$. The denoised auto-power is produced using the workflow shown in Figure \ref{fig:TASC_flowchart} and uses the TASC-based coherence calculated with a downsample factor of 6 as shown in Figure \ref{fig:Figure_downsample_factor_scan}. The increase in SNR is shown to be equivalent to that obtained using dual-channel spectral decorrelation CECE at $\rho_{\rm pol} = 0.976$. Narrowband peaks are the result of electronics noise.}
    \label{fig:Figure_denoised_autopower}
\end{figure*}

The magnitude of the TASC-based coherence spectrum shown in Figure \ref{fig:Figure_downsample_factor_scan} for a downsample factor of 6 was then multiplied by the auto-power spectral density calculated from the original time series to produce a denoised auto-power spectral density, per the workflow illustrated in Figure \ref{fig:TASC_flowchart}. As shown in Figure \ref{fig:Figure_denoised_autopower}, an increase in the SNR of about two orders of magnitude is observed in the denoised auto-power spectral density compared to the original auto-power spectral density. For the purpose of comparison, the cross-power spectral density was calculated using correlations between CECE channel \#19 and a radially-adjacent CECE channel (\#18). The localization ($\rho_{\rm pol} = 0.976$) of the cross-power spectral density is taken to be the mean of the two CECE channel measurements, which is very close to the radial localization ($\rho_{\rm pol} = 0.980$) of the channel \#19 measurement. The denoised auto-power spectral density is observed to be well-aligned with the cross-power spectral density over the entire range of frequencies shown, indicating that the increase in SNR obtained using the single-channel method is equivalent to that from dual-channel CECE. The results shown in Figures \ref{fig:Figure_downsample_factor_scan} and \ref{fig:Figure_denoised_autopower} validate the TASC method for a case of large-amplitude turbulent electron temperature fluctuations in the edge of an I-mode plasma. Calculating the electron temperature fluctuation amplitude ($\delta T_e / T_e$) using Eqn. 1 in Ref. \citenum{molina_cabrera_isotope_2023} for the TASC-based coherence (downsampled by 6) shown in Figure \ref{fig:Figure_downsample_factor_scan} using an integration region of 10 - 250 kHz and a background subtraction region of 250 - 300 kHz yields $\delta T_e / T_e = 4.6\%$, which confirms the large-amplitude nature of these fluctuations.

\begin{figure}
    \centering
    \begin{subfigure}[b]{0.48\textwidth}
         \centering
         \includegraphics[width=\textwidth]{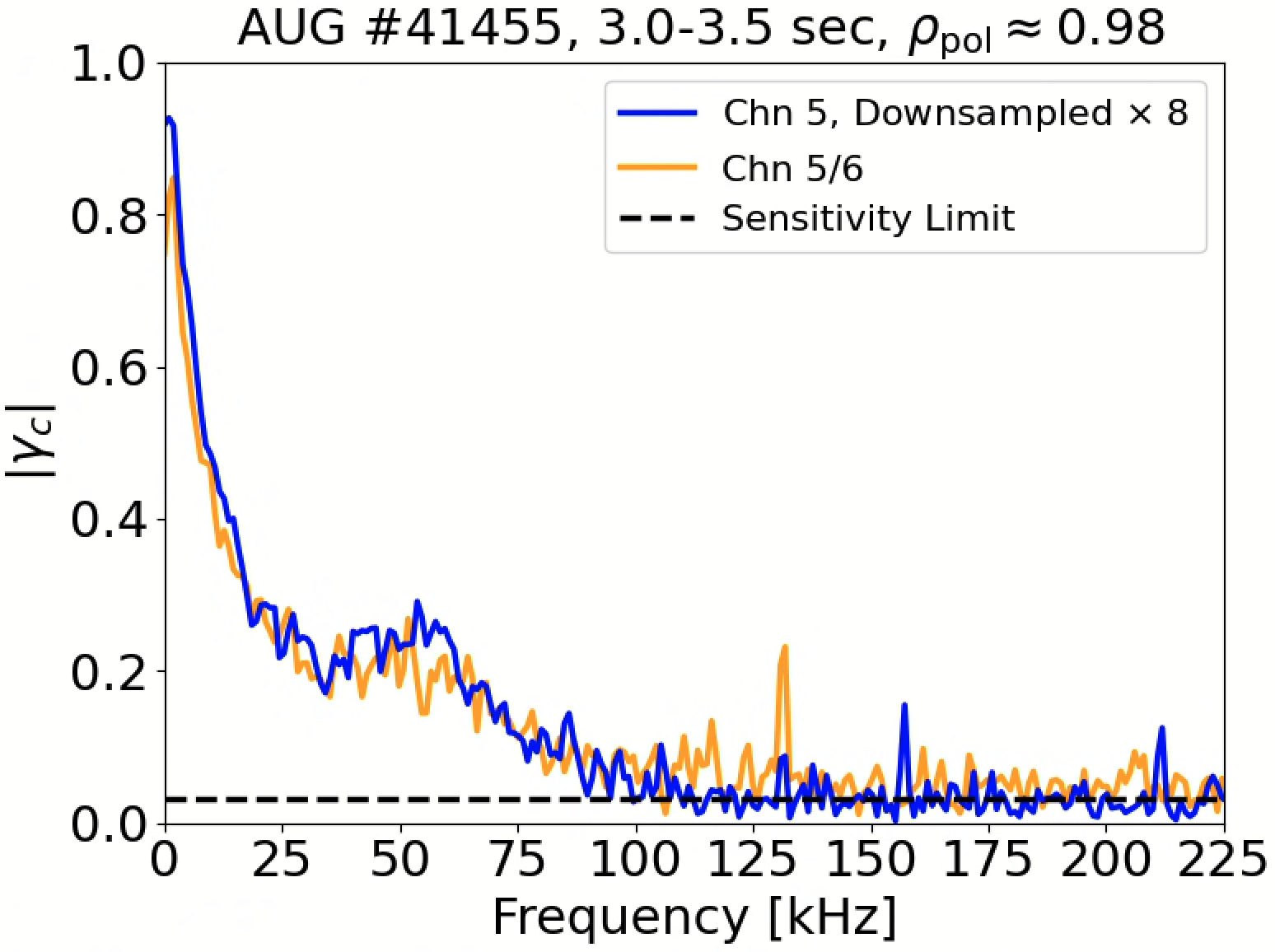}
    \end{subfigure}
    \caption{The magnitude of the TASC-based coherence spectra of the QCM compared to the dual-channel CECE spectrum in a small-ELM H-mode at $\rho_{\rm pol} \approx 0.98$. Narrowband peaks are the result of electronics noise.}
    \label{fig:Figure_Hmode_coherence}
    
\end{figure}

\begin{figure}
    \centering
    \begin{subfigure}[b]{0.48\textwidth}
         \centering
         \includegraphics[width=\textwidth]{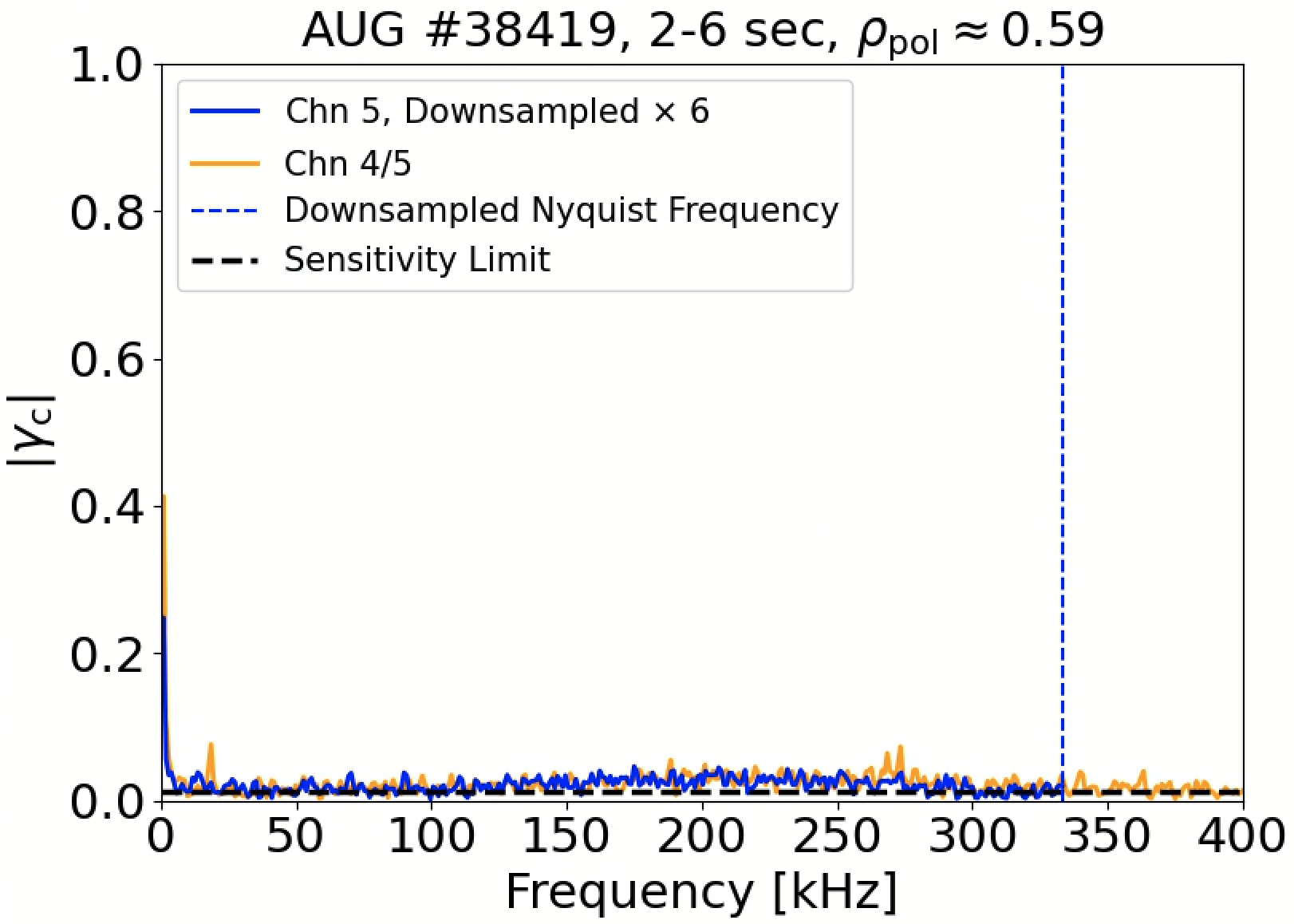}

    \caption{}
    \end{subfigure}
    
    \hfill
    \centering
    \begin{subfigure}[b]{0.48\textwidth}
         \centering
         \includegraphics[width=\textwidth]{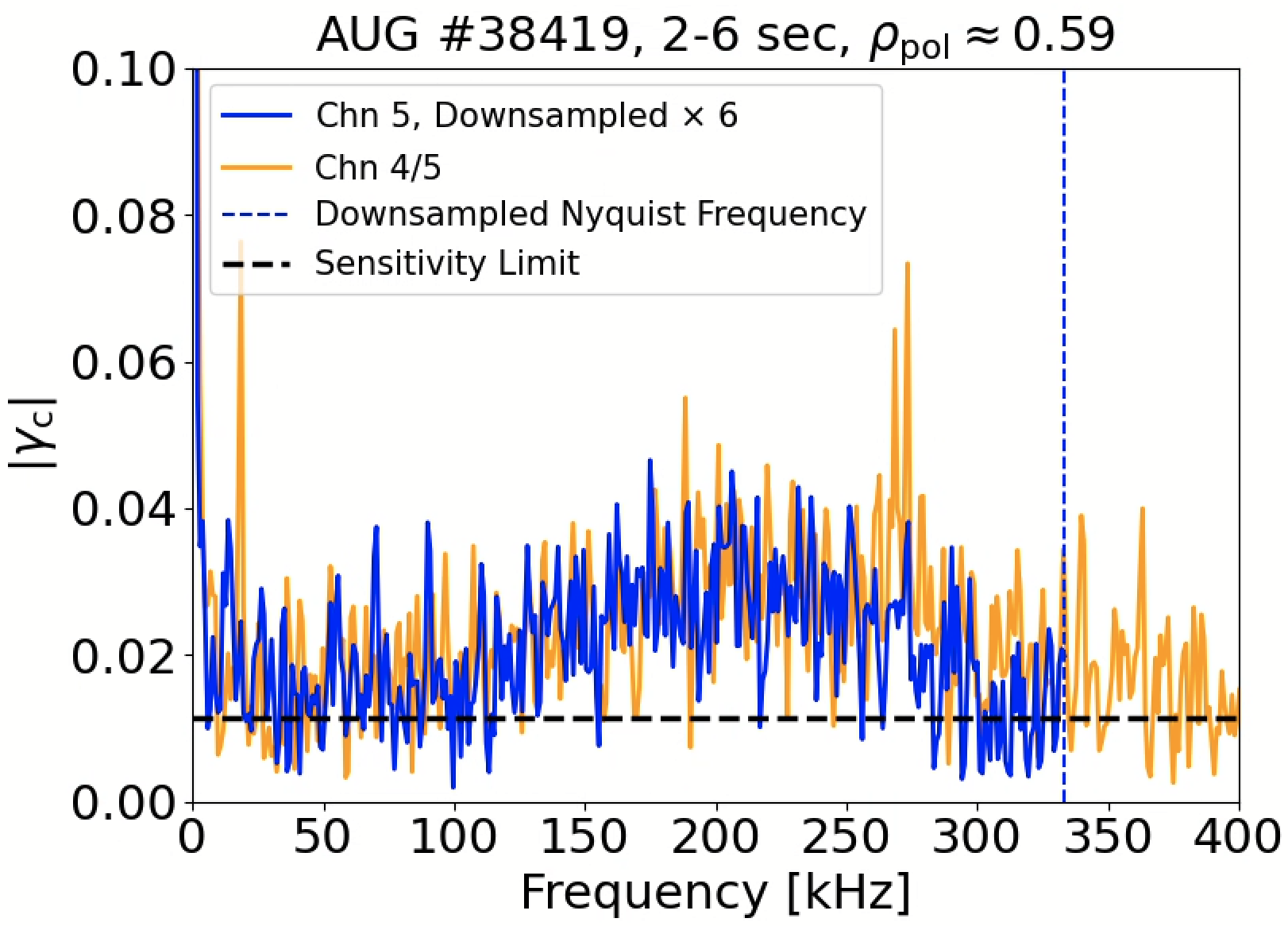}

    \caption{}
    \end{subfigure}
    \caption{The magnitudes of the TASC-based and dual-channel coherence spectra of broadband, low-amplitude L-mode core turbulence at $\rho_{\rm pol} \approx 0.59$. Narrowband peaks are the result of electronics noise. (a) y-axis shown up to 1 to emphasize the low coherence of the turbulent fluctuations. (b) Zoomed-in on the low-coherence signal.}
    \label{fig:Figure_Lmode_coherence}
    
\end{figure}

Additional validation of the TASC method is provided by Figures \ref{fig:Figure_Hmode_coherence} and \ref{fig:Figure_Lmode_coherence} for measurements of smaller-amplitude turbulent electron temperature fluctuations. Note that digital low-pass filtering of the time series data was not applied in either of these cases. Figure \ref{fig:Figure_Hmode_coherence} shows a case of fluctuations associated with the Quasi-Coherent Mode (QCM) in the edge of a high confinement (H-)mode plasma featuring small Edge Localized Modes (ELMs) from AUG discharge 41455, which is described in Ref. \citenum{yoo_design_2025}. The single-channel CECE coherence spectrum produced by downsampling by a factor of 8 closely matches the dual-channel CECE spectrum evaluated using two radially-adjacent CECE channels. Integrating the single-channel ECE coherence spectrum over 20 - 110 kHz and applying a background subtraction region of 150 - 200 kHz yields $\delta T_e / T_e = 1.38\%$, approximately three times smaller than the I-mode WCM peak fluctuation amplitude.

Figure \ref{fig:Figure_Lmode_coherence} shows a case of low-amplitude, broadband electron temperature fluctuations in the core of a low confinement (L-)mode plasma from AUG discharge 38419, which is described in Refs. \citenum{hofler_phd_thesis} and \citenum{hofler_milestone_2025}. The single-channel CECE coherence spectrum produced by downsampling by a factor of 6 closely matches the dual-channel CECE spectrum up to the downsampled Nyquist frequency of 333.3 kHz. Integrating the single-channel CECE coherence spectrum over 10 - 300 kHz and applying a background subtraction region of 300 - 325 kHz yields $\delta T_e / T_e = 0.75\%$, which is significantly lower in amplitude than the I-mode and H-mode edge fluctuations and also much broader in frequency. Therefore, this L-mode fluctuation measurement serves as a stringent validation of the TASC method as a means to accurately measure low-amplitude, broadband turbulent electron temperature fluctuations using individual radiometer channels.

\section{Discussion}
\label{sec:discussion}

The application of the TASC method to the analysis of electron temperature fluctuations measured using the CECE diagnostic in the core and edge of L, I, and H-mode plasmas shows that this method can accurately increase the SNR of broadband turbulence to the same degree as the dual-channel spectral decorrelation method. As already noted, the accuracy of the TASC method depends on the minimization of broadband, non-thermal electronics noise up to the digitizer Nyquist frequency to reduce the aliasing of this noise during the downsampling process. Another limitation of the TASC method is that it requires sufficiently high video bandwidths and digitizer sample rates to meet the criteria specified in Section \ref{sec:methods}.

It is important to note that the TASC method cannot be used to directly calculate correlation lengths since that requires the cross-correlation of spatially-separated radiometer channels. However, designing a CECE diagnostic for use with the TASC method does not preclude correlation length measurements, since this analysis can be done in conjunction with standard dual-channel methods. Furthermore, the TASC method could enable radiometer channels to be spaced further apart than in conventional CECE diagnostics while still retaining multiple channels within a turbulence correlation length, thereby allowing calculations of correlation lengths using standard cross-correlation techniques while covering a larger total plasma volume.

\section{Conclusion}
\label{sec:Conclusion}

This paper presents a new analysis method to reduce thermal noise and facilitate the measurement of broadband, low-amplitude electron temperature fluctuations using individual radiometer channels. This method takes advantage of differences in the correlation time of thermal noise compared to the correlation time of plasma fluctuations. A validation of this method was carried out using measurements of core and edge electron temperature fluctuations in L, I, and H-mode plasmas at AUG. This method opens up new possibilities for radiometer measurements of low-amplitude signals and, in addition, may be of use more generally in applications where small-amplitude signals are buried in white noise. Further work is required to identify and reduce sources of broadband, high frequency electronics noise in the AUG CECE diagnostic to prevent aliasing at lower frequencies during the single-channel method's downsampling process.

\section*{Acknowledgments}

This work is supported by the US Department of Energy under grants DE-SC0014264, DE-SC0006419, and DE-SC0017381. This work has been carried out within the framework of the EUROfusion Consortium, funded by the European Union via the Euratom Research and Training Programme (Grant Agreement No 101052200 — EUROfusion). Views and opinions expressed are however those of the author(s) only and do not necessarily reflect those of the European Union or the European Commission. Neither the European Union nor the European Commission can be held responsible for them. This material is based upon work supported by the National Science Foundation Graduate Research Fellowship under Grant No. 1745302 and 2141064. Any opinion, findings, and conclusions or recommendations expressed in this material are those of the authors and do not necessarily reflect the views of the National Science Foundation.

\section*{Author Declarations}

\subsection*{Conflicts of Interest}

The authors have no conflicts to disclose.

\subsection*{Author Contributions}

\textbf{C. Yoo}: Conceptualization (lead); Data curation (lead); Formal analysis (lead); Investigation (lead); Methodology (lead); Software (lead); Validation (lead); Visualization (lead); Writing - original draft (lead); Writing - review and editing (lead). \textbf{G. D. Conway}: Investigation (supporting); Methodology (supporting); Supervision (equal); Writing - review and editing (equal). \textbf{J. Schellpfeffer}: Validation (equal). \textbf{R. Bielajew}: Investigation (supporting). \textbf{K. H\"ofler}: Investigation (supporting). \textbf{D. J. Cruz-Zabala}: Investigation (supporting). \textbf{D. Cusick}: Investigation (supporting); Writing - review and editing (supporting). \textbf{W. Burke}: Validation (supporting). \textbf{B. Vanovac}: Investigation (supporting). \textbf{A. E. White}: Funding acquisition (lead); Methodology (supporting);  Project administration (lead); Resources (lead); Supervision (lead); Visualization (supporting); Writing - review and editing (equal).

\section*{Data Availability}

The data that support the findings of this study are available from the corresponding author upon reasonable request.

\section*{Appendix: example of the impact of aliased, broadband, high frequency electronics noise}
\label{sec: Appendix}

This section provides additional detail on the importance of reducing broadband, high frequency electronics noise in order to produce accurate spectra using the TASC-based, single-channel CECE analysis method. As noted in Section \ref{sec:results}, the time series data used to produce the single-channel spectra shown in Figures \ref{fig:Figure_downsample_factor_scan} and \ref{fig:Figure_denoised_autopower} was digitally low-pass filtered at 900 kHz. The purpose of applying this digital filter was to reduce the aliasing of broadband, high frequency
electronics noise during downsampling.

\begin{figure}
    \centering
    \begin{subfigure}[b]{0.48\textwidth}
         \centering
         \includegraphics[width=\textwidth]{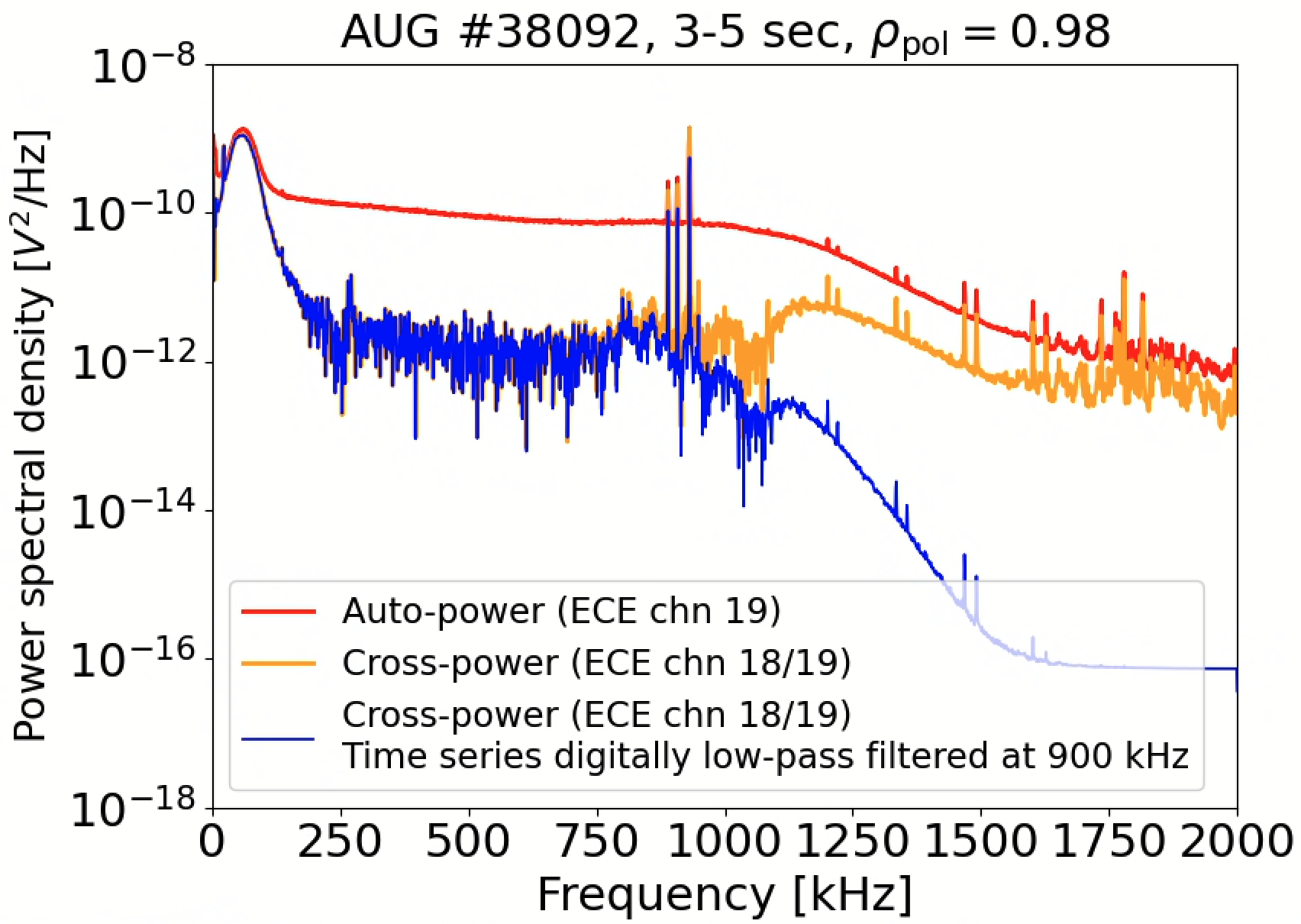} 
    \end{subfigure}
    \caption{The auto-power spectrum (red) exhibits a roll-off due to the hardware-based low-pass filter with a 1 MHz cutoff frequency. However, the dual-channel cross-power spectrum (gold) shows broadband, electronics noise in excess of 1 MHz that is correlated between the two channels. Digitally low-pass filtering the time series at 900 kHz reduces this electronics noise in the resulting cross-power spectrum (blue).}
    \label{fig:Figure_filtered_crosspower}
\end{figure}

Figure \ref{fig:Figure_filtered_crosspower} shows the auto-power spectrum (red) of CECE channel \#19, which exhibits a gradual roll-off above 1 MHz due to the hardware-based low-pass filter (4-pole Butterworth with a 1 MHz cutoff frequency). However, the dual-channel cross-power spectrum (gold) between radially-adjacent CECE channels \#18 and \#19 features significant broadband electronics noise in excess of 1 MHz that is correlated between the two channels and is not clearly visible in the channel \#19 auto-power spectrum. By first digitally low-pass filtering the time series data (using a 4-pole, Butterworth low-pass filter), this broadband electronics noise is significantly reduced in the resulting cross-power spectrum (blue). 900 kHz was chosen as the cutoff frequency for the digital filter in order to achieve sufficient attenuation by 1 MHz to reduce the high frequency noise while simultaneously being high enough in frequency to avoid a significant increase in the correlation time of the thermal noise. The TASC-based, single-channel spectra produced using this digitally low-pass filtered time series data is shown in Figures \ref{fig:Figure_downsample_factor_scan} and \ref{fig:Figure_denoised_autopower}.

\begin{figure}
    \centering
    \begin{subfigure}[b]{0.48\textwidth}
         \centering
         \includegraphics[width=\textwidth]{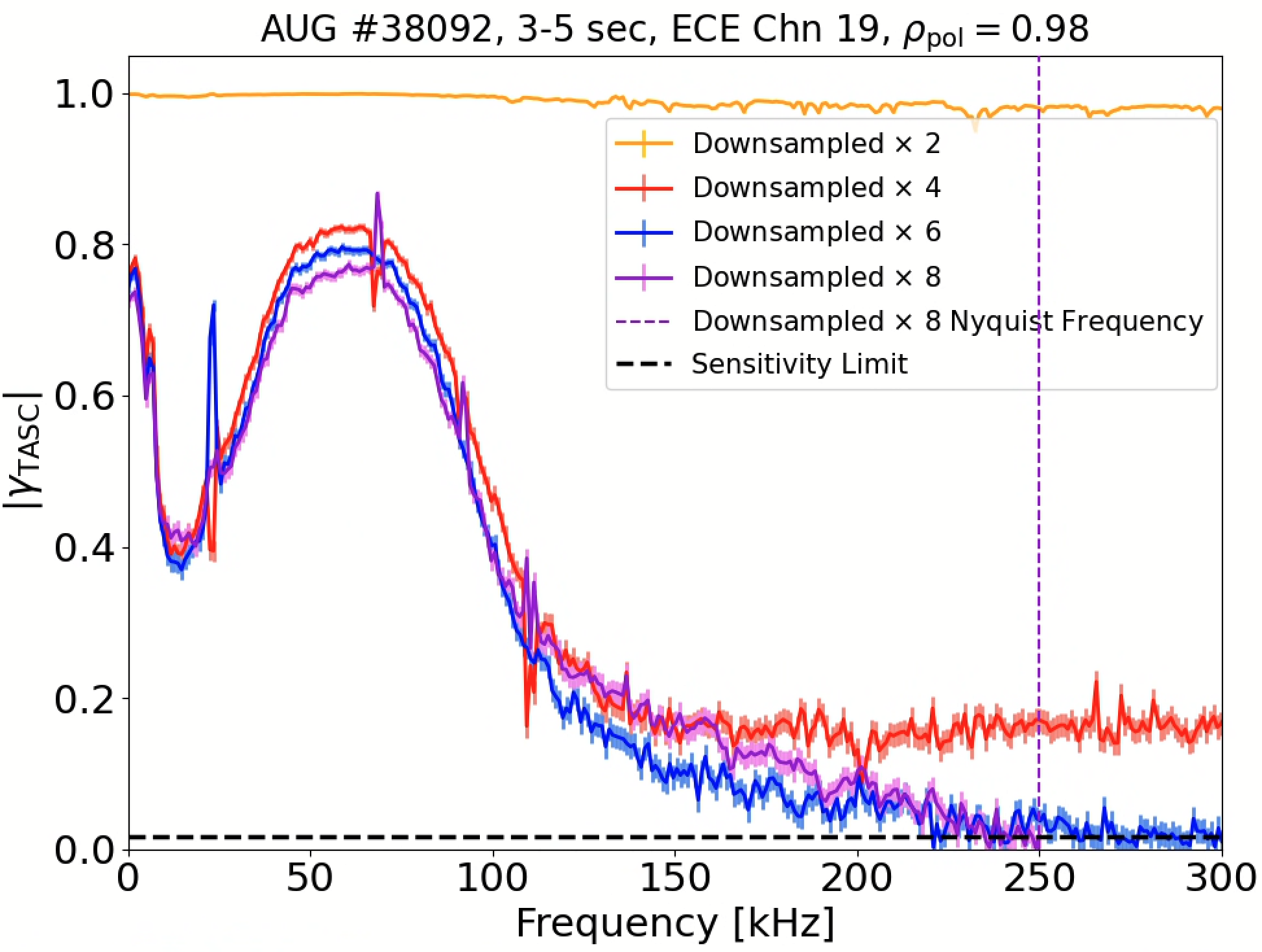} 
    \end{subfigure}
    \caption{Coherence spectra produced when the time series data corresponding to spectra shown in Figures \ref{fig:Figure_downsample_factor_scan}, \ref{fig:Figure_denoised_autopower}, \ref{fig:Figure_filtered_crosspower} is not first digitally low-pass filtered at 900 kHz. The downsampled $\times$ 8 spectrum is erroneously greater in magnitude than the downsampled $\times$ 6 spectrum at various frequencies as a result of the broadband electronics noise above 1 MHz shown in Figure \ref{fig:Figure_filtered_crosspower}.}
    \label{fig:Figure_filtered_downsample_factor_scan}
\end{figure}

If the time series data is not first digitally low-pass filtered, the TASC-based, single-channel spectra contain erroneous features as a result of the aliasing of this broadband, high frequency noise. The resulting single-channel coherence spectra are shown in Figure \ref{fig:Figure_filtered_downsample_factor_scan}. In particular, the downsampled $\times$ 8 coherence spectrum is erroneously greater in magnitude than the downsampled $\times$ 6 spectrum at frequencies above 100 kHz, as well as below 20 kHz, as a result of the broadband, high frequency electronics noise above 1 MHz shown in Figure \ref{fig:Figure_filtered_crosspower}. While digitally low-pass filtering at 900 kHz is clearly beneficial in this case, it is important to avoid digitally low-pass filtering too close to the downsampled Nyquist frequency, as noted in Section \ref{sec:methods}.

\section*{References}
\bibliographystyle{unsrt}
\bibliography{references}

\end{document}